\documentclass[runningheads]{llncs}

\usepackage[hyphens]{url}
\usepackage[hidelinks]{hyperref}
\usepackage[utf8]{inputenc}
\usepackage{subcaption}
\usepackage{mathabx}
\usepackage{microtype}
\usepackage{multirow}
\usepackage[normalem]{ulem}

\newcommand\Small{\fontsize{9}{9.2}\selectfont}
\newcommand*\LSTfont{\Small\ttfamily\SetTracking{encoding=*}{-60}\lsstyle}

\usepackage{listings}
\usepackage{tikz}

\begin{document}

\title{Interactively Constructing Knowledge Graphs from Messy User-Generated Spreadsheets}
\titlerunning{Interactively Constructing Knowledge Graphs}
\authorrunning{M. Schröder et al.}

\author{
	Markus Schröder \and
	Christian Jilek \and
	Michael Schulze \and
	Andreas Dengel
}
\institute{
	Smart Data \& Knowledge Services Dept., DFKI GmbH, Kaiserslautern, Germany\\ \and
	Computer Science Dept., TU Kaiserslautern, Germany\\
	\email{\{markus.schroeder, christian.jilek, michael.schulze, andreas.dengel\}@dfki.de}
}

\maketitle

\begin{abstract}

When spreadsheets are filled freely by knowledge workers, they can contain rather unstructured content.
For humans and especially machines it becomes difficult to interpret such data properly.
Therefore, spreadsheets are often converted to a more explicit, formal and structured form, for example, to a knowledge graph.
However, if a data maintenance strategy has been missing and user-generated data becomes ``messy'', the construction of knowledge graphs will be a challenging task.
In this paper, we catalog several of those challenges and propose an interactive approach to solve them.
Our approach includes a graphical user interface which enables knowledge engineers to bulk-annotate spreadsheet cells with extracted information.
Based on the cells' annotations a knowledge graph is ultimately formed.
Using five spreadsheets from an industrial scenario, we built a 25k-triple graph during our evaluation.
We compared our method with the state-of-the-art RDF Mapping Language (RML) attempt.
The comparison highlights contributions of our approach.

\end{abstract}

\section{Introduction}
\label{sec:intro}

Spreadsheets are widely used by knowledge workers, especially in the industrial sector.
Their methodology enables a well understood, easy and fast possibility to enter data.
Table structures give an intuitive overview and let users sort and filter data conveniently.
To capture data, a two dimensional cell matrix per sheet is used.
Yet, this structure does not predetermine how cells should be filled by users. 
As a consequence, cells can contain unstructured content which can be arbitrarily arranged in the sheet.
Because of that it could become rather difficult for humans and especially for machines to interpret such data properly.
Therefore, spreadsheets are often converted to a more explicit, formal and structured form.
A common practice is to transform spreadsheet contents to a connected graph of data -- a knowledge graph (KG) \cite{hogan2020knowledge}.
In this process, it is usual to consult external data sources containing structured information (like named entities) to better comprehend unstructured content.
However, not always do appropriate sources exist which forces knowledge engineers to consult human experts and explore the data interactively \cite{DBLP:conf/esws/Schroder19}.

Because spreadsheets do not enforce any modelling constraints, users are able to enter data freely.
Without a data maintenance strategy incorporated in their everyday work, this user-generated data can easily become inconsistent, heterogeneous and incomprehensible -- in short, ``messy''. Similarly, this is also recognized by Assem et al. \cite{ConvertingAndAnnotatingQuantitativeDataTables} reporting about researchers who tend to fill spreadsheets in a ``sloppy'' way.
Once such data shall be enriched to construct a complete and correct KG, several challenges have to be solved.

While working with messy data, we discovered typical challenges we had to overcome repeatedly.
To catalog these challenges, \autoref{table:example} outlines an exemplary spreadsheet that is conceived in a way to demonstrate them.
This small sheet describes documents (lines 1, 2 and 4) and their attachments (line 3) together with their associated departments, editors, revision types, change histories, published dates and sent status.
\begin{table}
	\centering
	\caption{
		An exemplary spreadsheet to demonstrate typical challenges when messy data has to be enriched.
	}
\begin{tabular}{|l|l|l|l|l|l|l|l|}
		\hline
		Line & Document ID & Dep. & Editor & Type & Changes & Published & Sent \\
		\hline
		1 & *AB-ztad.63/23 & GA & \sout{Cooper} Smith & C & V1: 2015-03-02 & 42415 & x\\
		\hline
		2 & AB-hzyx-78/24 & GA/BZ & Emma Thomas & N &  & TODO &  \\
		\hline
		3 & AB-hzyx-78/24 A1 & GA/BZ & Smith, Leo &  &  &  &  \\
		\hline
		4 & AB 5-pbga.67 & BZ & (new) Smith & ed.c & V1: Dec2009 & 15.05.2010 & - \\
		  &            &     & Thomas, E. &      & V2: Mar2010 &  & \\
		\hline
	\end{tabular}
	\label{table:example}
\end{table}

\noindent
In the following we discuss several challenges that inevitably occur when such a spreadsheet is interpreted.
\begin{description}

\item[Multiple Surface Forms] 
	 	A surface form is a text which is used to refer to an entity.
		Entities can be mentioned in various ways as demonstrated in the \textit{Editor} column for person entities.
		For instance, the editor Mrs. Thomas has two surface forms: ``Emma Thomas'' and ``Thomas, E.''. 
Thus, an extraction algorithm should be able to discover entities of different surface forms and reconciliate them properly, like in \cite{NamedEntityNormalizationInUserGeneratedContent}.
		Because algorithms are usually not flawless, a human expert should be able to correct errors.

\item[Mixed Date Representations] 
		Dates can be represented in various formats.
		In our example four formats can be found: \verb|YYYY-MM-DD|, \verb|MonthYYYY|, \verb|DD.MM.YYYY| and days since 1970 epoch.
		Note that the latter is usually represented in spreadsheets with a number while the other formats are texts.
A robust algorithm should be able to detect all kinds of date representations and convert them to a unified form.
	
\item[Acronyms and Symbols] 
	To reduce typing effort, users tend to write acronyms or even single symbols.
	That is why the words \underline{C}hanged, \underline{N}ew, and \underline{ed}itorial \underline{c}hange are abbreviated in the \textit{Type} column.
	An `x' symbol in the \textit{Sent} column is used to indicate that the document was sent while a `-' (minus) symbol or empty cell indicates the opposite.
Such insights should be integrated during the processing to extract such facts and entities correctly.

\item[Free Comments] 
	In spreadsheets, users are able to edit cells in order to write small comments.
	In our scenario, a document is marked with a preceding asterisk symbol (*) before the document ID or ``(new)'' is written in front of a last name to indicate, that the person was added recently.
Such short comments can easily become a distraction when named entities are extracted or cell contents are compared.
	
\item[Style Usage] 
		Spreadsheets allow the usage of different font styles, border types and colors.
		Changing style is a common way to express additional information in cells.
		In our example, Cooper is no longer an editor for a document:
		because of traceability reasons the name is not removed completely, but instead has been simply struck out.
If such style usage is not considered, the corresponding information will be lost. 
	
\item[Multiple Entities in a Cell]
		If a table is unnormalized, data redundancies can occur.
		A consequence is that spreadsheet cells contain multiple entities, like several persons in the \textit{Editor} column and more than one entry in the \textit{Changes} column.
Thus, cell contents have to be split appropriately to retrieve a set of entities instead of only a single one.
	
\item[Multiple Types in a Table] 
		If entity types share some properties (expressed with identical columns), they are sometimes listed in the same spreadsheet table.
		In our example, there are actually two types: documents and attachments.
		To distinguish them, an attachment is recognized whenever an `A' is followed by a number at the end of the document ID.
This and similar situations have to be considered when dynamically assigning types to extracted row entities.
	
\item[Implicit Relationship]
		Usual spreadsheets do not allow to explicitly state relationships among cells or rows (like foreign keys in databases).
		Thus, users tend to write identical texts (like IDs) in different cells to express an implicit relationship.  
		In our example, in order to know which attachment belongs to which document, the user repeated in Line 3 the document ID \verb|AB-hzyx-78/24| and appended \verb|A1| to form the attachment ID.
Such relationships have to be discovered and made explicit.

\end{description}
Despite the identified challenges which come along with such data, we intend to construct a knowledge graph (KG) that reflects the spreadsheet's information completely and correctly.
Given \autoref{table:example}, the envisioned KG would consist of three documents, one document having an attachment, three departments, three persons, three types, three change entries, two published dates and one sent statement.
These typed resources would have literal properties based on the cells' contents and would be interlinked according to the table structure and implicit relationships.
For the statement about Cooper being an editor of the first document, a special property would be used to distinguish that his name is struck out.
Having such knowledge graphs, instead of spreadsheets, would enable highly sophisticated semantic services to assist knowledge workers in their daily work, for instance by introducing managed forgetting \cite{DBLP:journals/ki/JilekRNMTDF19}.

The remaining part of this paper is structured as follows:
In Section \ref{sec:relwork} we will discuss approaches in literature that build knowledge graphs, especially from spreadsheet data.
This is followed by our own approach in Section \ref{sec:approach}.
For evaluation (Sec. \ref{sec:eval}), we compare our method with a typical mapping approach.
Finally, we conclude the paper in Section \ref{sec:concl}.

\section{Related Work}
\label{sec:relwork}

Several approaches in literature annotate table-like structures to make their content more interpretable.
On the one hand, this includes the annotation of column names which is also known as semantic labeling \cite{DBLP:conf/semweb/PhamAKS16}.
On the other hand, table contents are linked to unambiguous entities by using named entity recognition (NER), for example.
In this process various data sources can be utilized, ranging from taxonomies \cite{DBLP:conf/er/WangWWZ12} and search engines \cite{DBLP:conf/edbt/QuerciniR13} to knowledge bases like DBpedia \cite{DBLP:conf/seco/MulwadFJ12}\cite{DBLP:journals/jidm/BernardoMS13}\cite{DBLP:conf/esws/CremaschiRSP19} and Wikitology \cite{Syed2010ExploitingAW}.
Primarily, their motivation is to enhance the table's interpretability. 
With few more steps a KG could be built as a secondary product. However, their approaches heavily rely on existing data sources which contain (usually common knowledge) named entities.
In our scenario, there are no such external sources that could be used, for example, to link mentioned persons to a \textit{Who-is-Who} database.

In contrast, mapping approaches focus more on data conversion.
The goal of such methods is to map input data to a knowledge representation in order to form a KG.
To represent the output graph, the Resource Description Framework (RDF) \cite{rdf} and Web Ontology Language (OWL) \cite{owl} are commonly used.
For instance, Any23\footnote{\url{https://any23.apache.org/}} (Anything To Triples) automatically converts structured data, especially CSV, to RDF.
The algorithm converts every cell to a triple, whereas each row represents a resource and columns serve as properties.
More sophisticated mapping approaches typically provide a mapping language or domain specific language (DSL) to let the knowledge engineer express how the input data is mapped to a graph structure.
For example, Spread2RDF\footnote{\url{https://github.com/marcelotto/spread2rdf/}} uses a Ruby-internal language while $M^2$ (Mapping Master) \cite{o2010m2} builds upon the
Manchester Syntax since it specialises in transformation to OWL.
Tarql\footnote{\url{https://tarql.github.io/}} extends the SPARQL Protocol and RDF Query Language (SPARQL) to allow the mapping of spreadsheets by using construct queries.
The RDF Mapping Language (RML) \cite{dimou2014rml}, a superset of the W3C-recommended mapping language R2RML \cite{r2rml}, uses RDF to encode the mapping rules.
Alternatively, YARRRML \cite{DBLP:conf/esws/HeyvaertMDV18}, a human-readable subset of YAML, can be used.
RML can be further extended with the Function Ontology\footnote{\url{https://fno.io/}} (F\textsubscript{n}O) \cite{DBLP:conf/esws/MeesterDVM16}  to call custom transformation functions during the mapping.

All mapping approaches above differ in their expressiveness and complexity.
They often require to learn a new language in order to be able to express how the mapping should be performed.
When it comes to messy input data, several exceptions in data structure and content have to be considered which 
can make methods insufficient or force knowledge engineers to formulate more complicated rules.

In order to let users visually see and edit mappings, interactive tools additionally provide graphical user interfaces (GUI). 
Mindswap's Excel2RDF\footnote{\url{https://web.archive.org/web/20080520234848/http:/www.mindswap.org/~rreck/excel2rdf.shtml}} and its successor Convert2RDF\footnote{\url{https://web.archive.org/web/20080327181331/http://www.mindswap.org/~mhgrove/convert/}} seem to be one of the first tools that provide a GUI to map spreadsheets to RDF, yet with very limited functionality.
Often, mapping approaches implement a simple front-end for rule editing and output review, like for instance in RML's Matey\footnote{\url{https://rml.io/yarrrml/matey/}}.
In contrast, OpenRefine\footnote{\url{https://openrefine.org/}} provides a rich set of capabilities to explore, clean and transform messy data.
Its RDF extension\footnote{\url{https://github.com/stkenny/grefine-rdf-extension/}} implements mapping functionalities to create RDF.
Similarly, the Sheet2RDF plugin from Fiorelli et al. \cite{DBLP:conf/ieaaie/FiorelliLPST15} allows to edit mappings graphically which are translated into ProjEction of Annotations Rules (PEARL).
Since all these methods use a similar skeletal transformation specification, they are comparable with the Tarql approach mentioned above.
With the goal of integrating various sources, the tool Karma \cite{DBLP:conf/esws/GuptaSKGTM12} is also able to produce RDF.

Interactive tools support knowledge engineers in mapping input data to an intended output in a graphical way.
However, the discussed tools (except OpenRefine) are not intended to handle messy data since they are usually designed to work with clean data only.
Although OpenRefine is intended for scenarios with messy data, the data cleansing and transformation pre-phase does not harmonize well with the RDF generation phase.

\section{Approach}
\label{sec:approach}

\begin{figure}
	\centering
	\includegraphics[width=\textwidth]{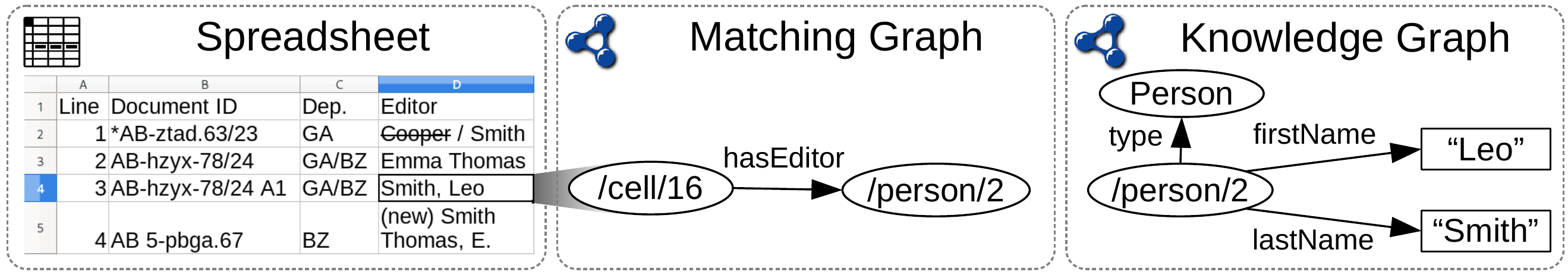}
	\caption{This demonstrates how a cell is annotated with a person resource in our approach:
			 The spreadsheet's cells (left) are assigned with deep links which are used in the matching graph (middle) to connect them with knowledge resources (right).
		     The URIs are shortened due to lack of space.
	}
	\label{fig:approach}
\end{figure}

Our interactive approach provides a graphical front-end for knowledge engineers (KEs).
Instead of defining a mapping with a mapping language, our approach lets users incrementally annotate spreadsheet cells with RDF statements.
\autoref{fig:approach} illustrates this annotation mechanism.
A view is provided (left) to let users browse loaded spreadsheets as in usual tools.
Unlike other tools, our program assigns to each spreadsheet cell a unique Uniform Resource Identifier (URI) using a deep linking approach \cite{SchroederJilekDengel2018} (e.g. \verb|/cell/16|).
This way we are able to make RDF statements about cells in a separated matching graph (middle). 
By this, we established an annotation mechanism using RDF.
Besides, our approach consists of various extraction methods (discussed below) that extract information from selected spreadsheet cells. 
The extracted information (like domain-specific resources) is stored in a separate RDF-based knowledge graph (right).
Spreadsheet cells can be easily associated with resources from the knowledge graph by looking up matchings in the matching graph.
By this we enrich the sheets' content with knowledge-related annotations.   
The made associations can be inspected and edited by KEs at any time during the process.

Our approach can be divided into two phases: in the first phase, spreadsheets are incrementally annotated using automated procedures invoked by the user.
We can further divide this phase into a staging and commit part.
During the staging, the KE selects cells, invokes a extraction procedure, reviews the procedure's outcome and carries out manual adjustments.
In the commit part, cell matches and acquired knowledge are persisted in the respective graphs.
Regarding the second phase, spreadsheet rows with annotated cells are interpreted as instances which are interlinked to finally form a knowledge graph.

In the following, our extraction methods are explained in detail.
To demonstrate their functionality, the exemplary spreadsheet in \autoref{table:example} is used.

\begin{figure}
	\centering
	\includegraphics[width=\textwidth]{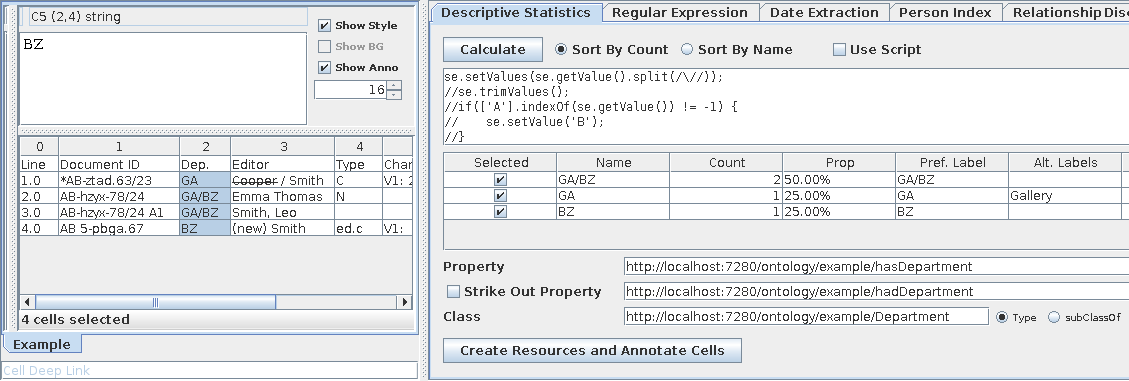}
	\caption{Four department cells are selected (left) in order to calculate a summary statistic yielding three departments (right).}
	\label{fig:ds}
\end{figure}

\noindent
\textbf{Descriptive Statistics.}
Whenever categorical data is analyzed, it is common to generate a summary statistic.
Such an overview reveals how often certain values occur in the data.
Using the same mechanism, we allow to discover potential knowledge resources mentioned in spreadsheet cells.
In this process, multiple equal values will be associated with a single resource in the knowledge graph.
Because cells could contain multiple entries or be improperly formatted, a JavaScript transformation script can be applied.
This way, the cell's content can be split (or transformed) in a desired manner.
\autoref{fig:ds} presents a screenshot showing how department names are discovered.
On the left side, the KE selects the relevant cells in the spreadsheet. 
On the right side, the descriptive statistic is calculated. 
Three departments are discovered and shown in the table.
For each row the user can decide on the resource's creation (using a checkbox) and adjust the resource's preferred label, alternative labels and comments.
Below, the property to annotate selected cells and the resources' type (or subclass) can be defined.

This and further methods are able to recognize struck out texts.
Once they discover such text parts, a different property in the matching graph is used to capture this information.

\begin{figure}
	\centering
	\includegraphics[width=\textwidth]{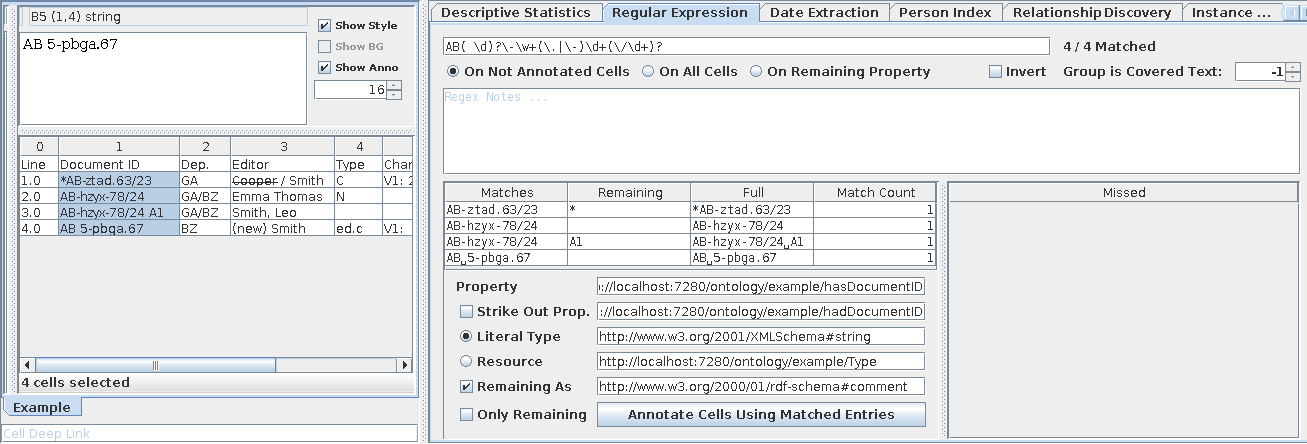}
	\caption{Document IDs are selected by the user (left) to extract only their ID part by applying a regular expression (right).}
	\label{fig:regex}
\end{figure}

\noindent
\textbf{Regular Expressions.}
Once spreadsheet cells contain regular syntax, a regular expression (regex) can be used to match or extract those parts.
For each application, we show matched and missed entries simultaneously to reduce trial and error rounds of pattern construction.  
Whenever a pattern matches, we have two possibilities: either, the cell is annotated with the extracted text as a literal value, or the cell is annotated with a constant resource.
In the former case, the user has to define the literal's type (e.g. string, int, etc). 
If there is remaining unmatched text left, an additional annotation can be made to store it.
Regex groups can be used to extract specific sections of the patterns. 
In contrast, the latter case can be used to assign class types to spreadsheet cells on matches.
\autoref{fig:regex} demonstrates the module's usage by extracting document IDs.
With the pattern in the text field, the four IDs can be found.
Remaining text like the asterisk (*) remark is captured as a comment.

\begin{figure}
	\centering
	\includegraphics[width=\textwidth]{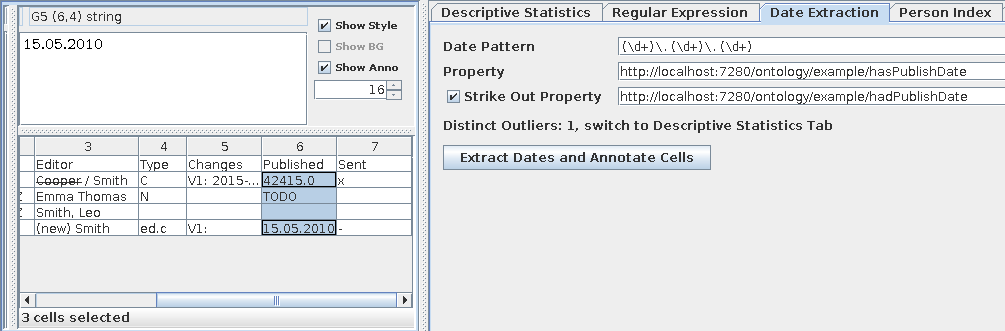}
	\caption{The dates (left) are extracted with a date pattern if the date is not a number (right).}
	\label{fig:date}
\end{figure}

\noindent
\textbf{Date Extraction.}
The extraction of dates is a special case of the previous one because in spreadsheets dates are usually represented as numbers that count the days since 1970 epoch.
In those cases it is easy to interpret dates properly.
However, if we encounter mixed date representations in texts, manually definable regular expressions are used as a fall back.
If the cell contains neither of that, outliers are assumed which are handled by the descriptive statistics module.
Again, a property can be stated that is used to attach the date on the cell as an annotation.
As an example, \autoref{fig:date} shows how the publication dates of documents are extracted.
Since ``TODO'' is not a date, it is passed to the Descriptive Statistics module.

\begin{figure}
	\centering
	\includegraphics[width=\textwidth]{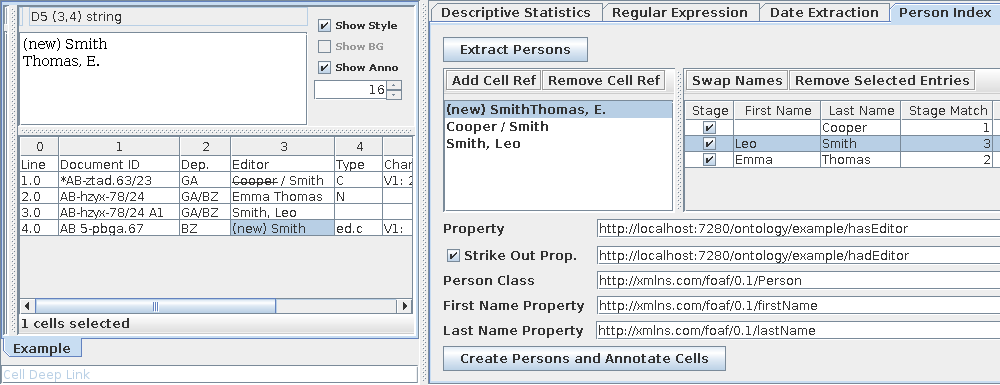}
	\caption{Editor names are first selected (left) in order to extract persons and build an person index (right).}
	\label{fig:person}
\end{figure}

\noindent
\textbf{Person Index.}
Beside dates, persons have to be extracted from texts, too.
Instead of simply listing discovered persons, we would like to build a person index which is a structured table that distinctly catalogs individuals by their names.
When persons are mentioned in texts with their first name and last name, there can be a high variation which of their names are used, how their names are ordered and if their names are abbreviated.
Therefore, it is difficult for unsupervised extraction algorithms to correctly find all persons.
A more detailed discussion of the problem can be found in a dedicated paper \cite{person-index}.
Here, we provide a user interface (\autoref{fig:person}) to allow the correction of possible errors made by the detection algorithm.
The rightmost table represents the person index.
If first name and last name have been extracted the wrong way, they can be swapped conveniently by the user.
When a KE selects a person, all the cells where the person is mentioned are shown in the left-sided list.
If cell references are missing or wrong, they can be added or removed, respectively.

\noindent
\textbf{Relationship Discovery.}
Cell contents can be filled in certain ways to express an implicit relationship among them.
First, our method uses two different regular expressions to divide selected cells into two groups.
Using pair-wise comparison between those groups, relationship conditions can be tested.
In our example,  the same ID prefix was used to identify which attachment belongs to which document.
Thus, we form a document and attachment group and check if the attachment ID starts with a document ID.

With the help of our \textbf{Inspection} view, all made annotations can be reviewed and possibly removed.
Once spreadsheet cells are selected, the associated knowledge is presented in an RDF serialization.

In the second phase, after all relevant cells are annotated using the previous methods, the KE finally constructs a knowledge graph using our \textbf{Instance Collector} module.
This method assumes that rows in a spreadsheet's table represent potential knowledge graph instances.
Hence, the table is scanned row-by-row for annotated cells to form instances.
A configurable default type is assigned if there is no type statement found in the scanning process.
Additionally, suggested instances can be filtered by enforcing required properties to be present.

\section{Evaluation}
\label{sec:eval}

To evaluate our approach, we use spreadsheet data from an industrial scenario.
Our exemplary sheet depicted in \autoref{table:example} was actually derived from that data.
In the following, the used dataset will be described only briefly for reasons of data protection. 
Then, the application of our approach is explained.
On the same data, the state-of-the-art RDF Mapping Language (RML) \cite{dimou2014rml} approach is applied.
RML was chosen because it is intuitive, provides a mapping tool and is well known and widely used in the Semantic Web community.
After matching our with RML's resulting graph instances, comparison of their properties highlight contributions of our approach.

To conduct the evaluation, a power supply company provided five spreadsheets:
four of them describe documents of different types, namely:
regulations, board decisions, directives and hazardous substance (haz.sub.) directives.
The fifth sheet is used to manage document revisions. \autoref{table:spreadsheets} lists the five sheets together with statistics about their content and cell types.
String cells which contain unstructured text are the most common ones followed by numeric cells mostly used to represent dates.
Since formulas are not relevant for the intended knowledge graph, we ignore them.
\begin{table}
	\centering
	\caption{
		Statistics about the five spreadsheets used in the evaluation. 
		For each sheet the number of rows, columns, non-empty cells, cell types and CSV file size are presented.
	}
\begin{tabular}{|l|r|r|r|r|r|r|r|}
		\hline
		Sheet      & Rows       & Columns & Cells & String & Numeric & Formular & CSV File Size \\
		\hline
		\hline
		Regulation     & 1218      & 30 & 21854 & 16451 & 3921   & 1198 & 360.0 kB\\
		\hline
		Board Decision & 57        & 22 & 525   & 472   & 53     & 0  & 7.5 kB  \\
		\hline
		Directive      & 126       & 21 & 1437   & 1244 & 193    & 0  & 21.0 kB  \\
		\hline
		Haz.Sub. Directive & 201  & 15 & 1174   & 913  & 261   & 0 &  22.0 kB \\
		\hline
		Revision            & 722  & 61 & 16105   & 7683 & 5528 & 2894 & 154.0 kB  \\
		\hline
		\hline
		Sum        & 2324       & 149   & 41095 & 26763 & 9956 & 4092 & 564.5 kB \\
		\hline
	\end{tabular}
	\label{table:spreadsheets}
\end{table}

Our approach was applied to the given data.
In discussions with domain experts we iteratively annotated spreadsheets until a ground truth was reached.
To handle data messiness in our approach properly, it is usual to have some manual effort for adjustments. 
\autoref{table:approach_result} lists for each spreadsheet how many columns were processed and how often certain methods were used.
Some columns were skipped because they are irrelevant for the intended KG.
Thus, appropriate cells were selected manually before the application of each method.
Regular expressions were used most often because the most common string cells contained regular syntax, like IDs or symbols.
For exceptions in the regular syntax, we extended the expressions accordingly.
The second most common method was date extraction, especially in revisions where dates were recorded when certain subtasks have been completed.
Descriptive statistics are uniformly used whenever the cells refer to entities.
However, we often had to write transformation scripts when cells do not clearly refer to one entity.
Whenever persons were discovered and extracted, we had manual effort to correct errors made by our detection algorithm.
For each of the annotated sheets, instance collection and -- where necessary -- relationship discovery is performed.
After applying 82 times a method (not counting instance collection and relationship discovery), our resulting knowledge graph consists of 25016 triples that form 2719 instances having 15 types.
\begin{table}
	\centering
	\caption{
		For each spreadsheet the table lists the number of columns \mbox{(processed/total)} and how often extraction procedures were applied.
	}
\begin{tabular}{|l|r|r|r|r|r|}
		\hline
		Sheet               & Columns  & Desc. Stat. & Regex & Date Extraction & Person Index \\
		\hline
		\hline
		Regulation          & 24/30     & 5 & 16 & 2 & 1  \\
		\hline
		Board Decision      & 14/22     & 1 & 11 & 1 & 1  \\
		\hline
		Directive           & 14/21     & 3 & 5 & 3 & 3 \\
		\hline
		Haz.Sub. Directive &  12/15    & 2 & 5 & 2 & 3 \\
		\hline
		Revision            & 18/61    & 3 & 1 & 12 & 2 \\
		\hline
		\hline
		Sum & 82/149 & 14 & 38 & 20 & 10 \\
		\hline
	\end{tabular}
	\label{table:approach_result}
\end{table}

On the same given data, we applied the state-of-the-art RDF Mapping Language (RML) approach \cite{dimou2014rml}
by using the RML Mapper tool\footnote{\url{https://github.com/RMLio/rmlmapper-java/}} (version 4.7.0).
Since the tool does not support Excel as input, we used the command-line tool \verb|ssconvert| from Gnumeric\footnote{\url{http://www.gnumeric.org/}} to convert the Excel file to one CSV file per sheet (using \verb|-S| argument).
Note that in this transformation step we unfortunately loose style information (e.g. struck out text). Because RML assumes that header names are distinct, we have edited some column names accordingly.
Additionally, initial rows were removed so the header line occurs on top.
In order to replicate the same behavior of our approach, five custom functions were implemented by using the Function Ontology (F\textsubscript{n}O) \cite{DBLP:conf/esws/MeesterDVM16}.
In the end, we wrote 29 subject maps, 102 predicate-object maps and used a function call 116 times (with \verb|fnml:functionValue|). 
RML's final knowledge graph consists of 22619 triples that form 2901 instances having 15 types.

Next, we will compare our KG with RML's KG in two stages.
First, with focus on instances, we check if the two graphs contain the same typed resources:
for each instance in our graph, the semantically equal counterpart in RML's graph is tried to be matched.
This way we make sure that graphs are equivalent on an instance level.
Second, we examine the instances' properties.
Because our approach could solve all proposed challenges, our KG contains several additional statements about instances.
By this, contributions are identified which are novel to RML's functional scope.

In the first stage, we match instances class-wise using unambiguous property values, like for example, IDs or labels.
\autoref{table:eval_result} compares our method's output (the expected result) with the RML mapping outcome (the actual result) with regard to instances. For analysis, RML's precision and recall metrics \cite{DBLP:conf/muc/Chinchor92} are calculated. The table shows that the RML output almost always retrieves more instances than we do.
This is due to preceding selections of spreadsheet cells by our knowledge engineer to fine-tune inputs for extraction procedures.
Thus, a lower precision of RML may be expected.
In case of the person type, our detection algorithm produces some errors that were manually cleaned, too.
Since few persons share the same last name, RML generated two instead of one person instances, leading to a recall slightly below $1.00$.
Since RML's recall is $1.00$ in the other cases, semantically equal counterparts for instances were always found in our graph.
\begin{table}
	\centering
	\caption{
		Comparison of instances (inst.) generated by our method (expected) versus the RML mapping (actual).
After a simple instance matching, RML's precision and recall are calculated.
	}

\begin{tabular}{|l|r|r|r|r|}
	\hline
	Type & Our Inst. & RML Inst. & RML's Precision & RML's Recall \\
	\hline
	\hline
	Work Status & 4 & 5 & 0.80 & 1.00  \\\hline
	Transfer Type & 3 & 4 & 0.75 & 1.00  \\\hline
	Category & 47 & 51 & 0.92 & 1.00 \\\hline
	Protection Need & 2 & 3 & 0.67 & 1.00  \\\hline
	Type of Treatment & 4 & 6 & 0.67 & 1.00 \\\hline
	Department & 75 & 81 & 0.93 & 1.00 \\\hline
	Newsletter & 235 & 237 & 0.99 & 1.00 \\\hline
	Person & 135 & 155 & 0.85 & 0.98 \\\hline
	Logo & 3 & 6 & 0.50 & 1.00  \\\hline
	Board Decision & 23 & 24 & 0.96 & 1.00 \\\hline
	Haz.Sub. Directive & 53 & 55 & 0.96 & 1.00 \\\hline
	Directive & 108 & 111 & 0.97 & 1.00 \\\hline
	Revision & 654 & 757 & 0.86 & 1.00  \\\hline
	Attachment & 698 & 737 & 0.95 & 1.00 \\\hline
	Regulation & 675 & 675 & 1.00 & 1.00 \\\hline
\end{tabular}

	\label{table:eval_result}
\end{table}

In the second stage, we focus on the instances' properties and identify four contributions of our approach.
Three of them can be allocated to identified challenges from the first section that were solved successfully.

\noindent
\textbf{Multiple Entries in a Cell.} 
In our spreadsheet data, 510 cells contain more than one entity.
These cells consist of 1081 instances ($39,7\%$ of all 2719 instances).
Since all of our extraction methods consider multi-valued cells, all instance properties were extracted.
Unfortunately, we could not reproduce this in RML.
In fact, the RML specification states that ``[a triple map] must have exactly \textit{one subject map} that specifies how to generate \textit{a subject for each row}(\dots) of the logical source'' (draft from 21st January 2020) \cite{rml-spec}.
To our state of research, multi-valued cells are currently not yet considered in RML's approach\footnote{We contacted a corresponding author and also stated a Stackoverflow question at \url{https://stackoverflow.com/q/61751174/8540029}.}.

\noindent
\textbf{Implicit Relationship.} 
In our scenario, documents and their attachments are related to each other if they share the same prefix in their IDs.
Our relationship discovery method was able to uncover these implicit relationships and made 698 statements between documents and attachments.
Although, RML provides a join mechanism to generate relationships among logical sources, we could not apply it to reproduce the same behavior, since cells are joined on equal contents (i.e. not on prefix matches).

\noindent
\textbf{Style Usage.} 
In our experiment, five columns contained struck out texts.
All four extraction procedures are able to recognize this style usage and made in sum 236 statements by using five additional properties.
Since RML Mapper does not yet support Excel files natively, style information will be lost if spreadsheets are converted to CSV or similar formats.

\noindent
\textbf{Traceability.} 
In our approach, we additionally kept 40103 statements in our matching graph to enable traceability.
This becomes crucial when we would like to enrich the data source with the generated knowledge graph or when debugging is needed.
In fact, our graphical user interface is able to present the data source together with their annotations.
By this, a knowledge engineer can directly inspect the outcome of extraction methods.

\section{Conclusion and Outlook}
\label{sec:concl}

In this paper, we explicitly addressed messy data in spreadsheets and listed several challenges that occur when it is mapped to a knowledge graph.
Related work revealed that existing approaches have no or rather inconvenient strategies to handle such data.
That is why we proposed our interactive approach which enables knowledge engineers to bulk-annotate spreadsheet cells with extracted information.
Using provided data, we evaluated our approach by building a knowledge graph and compared it to a graph created by the RDF Mapping Language (RML) method.
Four contributions of our approach were identified: 
handling of multi-valued cells, discovery of implicit relationships, consideration of style usage and traceability.

In the future we aim to show that our approach generalizes to other (messy) datasets and that it is easy to be use by knowledge engineers.
In parallel, we plan to efficiently integrate domain experts in the building process as humans-in-the-loop.
This way, we would like to incorporate expertise and feedback the moment it becomes necessary.
Last but not least, our method is planned to be implemented as a module in our larger toolkit for building knowledge graphs in corporate memories \cite{Maus2013}.

\bibliographystyle{llncs}
\bibliography{paper}

\begin{thebibliography}{10}
\providecommand{\url}[1]{\texttt{#1}}
\providecommand{\urlprefix}{URL }
\providecommand{\doi}[1]{https://doi.org/#1}

\bibitem{ConvertingAndAnnotatingQuantitativeDataTables}
van Assem, M., Rijgersberg, H., Wigham, M., Top, J.L.: Converting and
  annotating quantitative data tables. In: The Semantic Web - {ISWC} 2010 - 9th
  International Semantic Web Conference, {ISWC} 2010, Shanghai, China, November
  7-11, 2010. Lecture Notes in Computer Science, vol.~6496, pp. 16--31.
  Springer (2010), \url{https://doi.org/10.1007/978-3-642-17746-0\_2}

\bibitem{DBLP:journals/jidm/BernardoMS13}
Bernardo, I.R., Mota, M.S., Santanch{\`{e}}, A.: Extracting and semantically
  integrating implicit schemas from multiple spreadsheets of biology based on
  the recognition of their nature. {JIDM}  \textbf{4}(2),  104--113 (2013)

\bibitem{DBLP:conf/muc/Chinchor92}
Chinchor, N.: {MUC-4} evaluation metrics. In: Proceedings of the 4th Conference
  on Message Understanding, {MUC} 1992, McLean, Virginia, USA, June 16-18,
  1992. pp. 22--29. {ACL} (1992), \url{https://doi.org/10.3115/1072064.1072067}

\bibitem{DBLP:conf/esws/CremaschiRSP19}
Cremaschi, M., Rula, A., Siano, A., Paoli, F.D.: Mantistable: {A} tool for
  creating semantic annotations on tabular data. In: The Semantic Web: {ESWC}
  2019 Satellite Events - {ESWC} 2019 Satellite Events, Portoro{\v{z}},
  Slovenia, June 2-6, 2019, Revised Selected Papers. Lecture Notes in Computer
  Science, vol. 11762, pp. 18--23. Springer (2019),
  \url{https://doi.org/10.1007/978-3-030-32327-1\_4}

\bibitem{rml-spec}
Dimou, A., Sande, M.V.: {RDF Mapping Language (RML)}.
  \url{https://rml.io/specs/rml/} (2020), accessed: 2020-05-01

\bibitem{dimou2014rml}
Dimou, A., Vander~Sande, M., Colpaert, P., Verborgh, R., Mannens, E., Van~de
  Walle, R.: {RML:} a generic language for integrated {RDF} mappings of
  heterogeneous data. In: Proceedings of the 7th Workshop on Linked Data on the
  Web. CEUR Workshop Proceedings, vol.~1184 (Apr 2014)

\bibitem{DBLP:conf/ieaaie/FiorelliLPST15}
Fiorelli, M., Lorenzetti, T., Pazienza, M.T., Stellato, A., Turbati, A.:
  Sheet2rdf: a flexible and dynamic spreadsheet import{\&}lifting framework for
  {RDF}. In: Current Approaches in Applied Artificial Intelligence - 28th
  International Conference on Industrial, Engineering and Other Applications of
  Applied Intelligent Systems, {IEA/AIE} 2015, Seoul, South Korea, June 10-12,
  2015, Proceedings. Lecture Notes in Computer Science, vol.~9101, pp.
  131--140. Springer (2015),
  \url{https://doi.org/10.1007/978-3-319-19066-2\_13}

\bibitem{DBLP:conf/esws/GuptaSKGTM12}
Gupta, S., Szekely, P.A., Knoblock, C.A., Goel, A., Taheriyan, M., Muslea, M.:
  Karma: {A} system for mapping structured sources into the semantic web. In:
  The Semantic Web: {ESWC} 2012 Satellite Events - {ESWC} 2012 Satellite
  Events, Heraklion, Crete, Greece, May 27-31, 2012. Revised Selected Papers.
  Lecture Notes in Computer Science, vol.~7540, pp. 430--434. Springer (2012),
  \url{https://doi.org/10.1007/978-3-662-46641-4\_40}

\bibitem{DBLP:conf/esws/HeyvaertMDV18}
Heyvaert, P., Meester, B.D., Dimou, A., Verborgh, R.: Declarative rules for
  linked data generation at your fingertips! In: The Semantic Web: {ESWC} 2018
  Satellite Events - {ESWC} 2018 Satellite Events, Heraklion, Crete, Greece,
  June 3-7, 2018, Revised Selected Papers. Lecture Notes in Computer Science,
  vol. 11155, pp. 213--217. Springer (2018),
  \url{https://doi.org/10.1007/978-3-319-98192-5\_40}

\bibitem{hogan2020knowledge}
Hogan, A., Blomqvist, E., Cochez, M., d'Amato, C., de~Melo, G., Gutierrez, C.,
  Gayo, J.E.L., Kirrane, S., Neumaier, S., Polleres, A., Navigli, R., Ngomo,
  A.C.N., Rashid, S.M., Rula, A., Schmelzeisen, L., Sequeda, J., Staab, S.,
  Zimmermann, A.: Knowledge graphs. arXiv 2003.02320  (2020)

\bibitem{NamedEntityNormalizationInUserGeneratedContent}
Jijkoun, V., Khalid, M.A., Marx, M., de~Rijke, M.: Named entity normalization
  in user generated content. In: Lopresti, D.P., Roy, S., Schulz, K.U.,
  Subramaniam, L.V. (eds.) Proceedings of the Second Workshop on Analytics for
  Noisy Unstructured Text Data, {AND} 2008, Singapore, July 24, 2008. {ACM}
  International Conference Proceeding Series, vol.~303, pp. 23--30. {ACM}
  (2008), \url{https://doi.org/10.1145/1390749.1390755}

\bibitem{DBLP:journals/ki/JilekRNMTDF19}
Jilek, C., Runge, Y., Nieder{\'{e}}e, C., Maus, H., Tempel, T., Dengel, A.,
  Frings, C.: Managed forgetting to support information management and
  knowledge work. {KI}  \textbf{33}(1),  45--55 (2019),
  \url{https://doi.org/10.1007/s13218-018-00568-9}

\bibitem{Maus2013}
Maus, H., Schwarz, S., Dengel, A.: Weaving Personal Knowledge Spaces into
  Office Applications, pp. 71--82. Springer (2013)

\bibitem{DBLP:conf/esws/MeesterDVM16}
Meester, B.D., Dimou, A., Verborgh, R., Mannens, E.: An ontology to
  semantically declare and describe functions. In: The Semantic Web - {ESWC}
  2016 Satellite Events, Heraklion, Crete, Greece, May 29 - June 2, 2016,
  Revised Selected Papers. Lecture Notes in Computer Science, vol.~9989, pp.
  46--49 (2016), \url{https://doi.org/10.1007/978-3-319-47602-5\_10}

\bibitem{DBLP:conf/seco/MulwadFJ12}
Mulwad, V., Finin, T., Joshi, A.: A domain independent framework for extracting
  linked semantic data from tables. In: Search Computing - Broadening Web
  Search, Lecture Notes in Computer Science, vol.~7538, pp. 16--33. Springer
  (2012), \url{https://doi.org/10.1007/978-3-642-34213-4\_2}

\bibitem{o2010m2}
O'Connor, M.J., Halaschek-Wiener, C., Musen, M.A.: {M2}: A language for mapping
  spreadsheets to {OWL}. OWL: Experiences and Directions (OWLED), Sixth
  International Workshop (2010)

\bibitem{DBLP:conf/semweb/PhamAKS16}
Pham, M., Alse, S., Knoblock, C.A., Szekely, P.A.: Semantic labeling: {A}
  domain-independent approach. In: The Semantic Web - {ISWC} 2016 - 15th
  International Semantic Web Conference, Kobe, Japan, October 17-21, 2016,
  Proceedings, Part {I}. Lecture Notes in Computer Science, vol.~9981, pp.
  446--462 (2016), \url{https://doi.org/10.1007/978-3-319-46523-4\_27}

\bibitem{DBLP:conf/edbt/QuerciniR13}
Quercini, G., Reynaud, C.: Entity discovery and annotation in tables. In: Joint
  2013 {EDBT/ICDT} Conferences, {EDBT} '13 Proceedings, Genoa, Italy, March
  18-22, 2013. pp. 693--704. {ACM} (2013),
  \url{https://doi.org/10.1145/2452376.2452457}

\bibitem{DBLP:conf/esws/Schroder19}
Schr{\"{o}}der, M.: Efficient high-level semantic enrichment of undocumented
  enterprise data. In: The Semantic Web: {ESWC} 2019 Satellite Events - {ESWC}
  2019 Satellite Events, Portoro{\v{z}}, Slovenia, June 2-6, 2019, Revised
  Selected Papers. Lecture Notes in Computer Science, vol. 11762, pp. 220--230.
  Springer (2019), \url{https://doi.org/10.1007/978-3-030-32327-1\_41}

\bibitem{SchroederJilekDengel2018}
Schröder, M., Jilek, C., Dengel, A.: Deep linking desktop resources. In: The
  Semantic Web: ESWC 2018 Satellite Events -- ESWC 2018 Satellite Events,
  Heraklion, Crete, Greece, June 3-7, 2018, Revised Selected Papers. pp.
  202--207. Springer (2018),
  \url{https://doi.org/10.1007/978-3-319-98192-5\_38}

\bibitem{person-index}
Schröder, M., Jilek, C., Schulze, M., Dengel, A.: The person index challenge:
  Extraction of persons from messy, short texts. (under review)  (2020)

\bibitem{Syed2010ExploitingAW}
Syed, Z., Finin, T.W., Mulwad, V., Joshi, A.: Exploiting a web of semantic data
  for interpreting tables. In: Proceedings of the Second Web Science Conference
  (2010)

\bibitem{DBLP:conf/er/WangWWZ12}
Wang, J., Wang, H., Wang, Z., Zhu, K.Q.: Understanding tables on the web. In:
  Conceptual Modeling - 31st International Conference {ER} 2012, Florence,
  Italy, October 15-18, 2012. Proceedings. Lecture Notes in Computer Science,
  vol.~7532, pp. 141--155. Springer (2012),
  \url{https://doi.org/10.1007/978-3-642-34002-4\_11}

\bibitem{owl}
{World Wide Web Consortium}: {OWL 2 Web Ontology Language Document Overview
  (Second Edition)}. \url{https://www.w3.org/TR/owl2-overview/} (2012),
  accessed: 2020-05-01

\bibitem{r2rml}
{World Wide Web Consortium}: {R2RML}: {RDB} to {RDF} mapping language.
  \url{https://www.w3.org/TR/r2rml/} (2012), accessed: 2020-05-01

\bibitem{rdf}
{World Wide Web Consortium}: {RDF 1.1 Primer}.
  \url{https://www.w3.org/TR/rdf11-primer/} (2014), accessed: 2020-05-01

\end{thebibliography}

\end{document}